\documentclass[preprint,prb]{revtex4}
\usepackage{natbib}
\bibpunct{}{}{,}{s}{,}{,}

\usepackage{graphicx}
\usepackage{epsfig}
\usepackage{ae}
\usepackage{aecompl}

\begin{document}
\bibliographystyle{apsrev}

\title{ Phase Separation of Saturated and Mono-unsaturated Lipids as
determined from a Microscopic Model}

\author{R. \ Elliott}
\author{K.\ Katsov}
\author{M. Schick}
\affiliation{Department of Physics, University of Washington,  Box
  351560, Seattle, WA 98195-1560 U. S. A.} 

\author{I. Szleifer}
\affiliation{ Department of Chemistry, Purdue University,
West Lafayette, IN 47907-1393 U. S. A.}

\date{\today, draft}

\begin{abstract} A molecular model is proposed of a bilayer consisting of
fully saturated DPPC and mono unsaturated DOPC. The model not only
encompasses the constant density within the hydrophobic core of the
bilayer, but also the tendency of chain segments to align.  It is solved
within self-consistent field theory.  A model bilayer of DPPC undergoes a
main chain transition to a gel phase, while a bilayer of DOPC does not do
so above zero degrees centigrade because of the double bond which disrupts
order. We examine structural and thermodynamic properties of these
membranes and find our results in reasonable accord with experiment. In
particular, order-parameter profiles are in good agreement with NMR
experiments.  A phase diagram is obtained for mixtures of these lipids in
a membrane at zero tension. The system undergoes phase separation below
the main-chain transition temperature of the saturated lipid. Extensions
to the ternary DPPC, DOPC, and cholesterol system are outlined.
\end{abstract}

\maketitle

\newpage \section {INTRODUCTION} The hypothesis that the lipids in the
plasma membrane are distributed inhomogeneously has received enormous
attention. Small domains, known as ``rafts", rich in saturated lipids and
cholesterol, have been implicated in many biological processes including
endocytosis, transcription and transduction processes, and viral
infection. The size and nature of such domains in biological membranes is
currently unclear \citep{edidin}. The situation in model membranes,
however, is more transparent.  Experimental studies in giant unilamellar
vesicles 
\citep{veatch2002,veatch2003, dietrich}  containing
mixtures of saturated and unsaturated lipids and cholesterol, which mimic
the components of the outer leaflet of the plasma membrane, show the
existence of at least three phases; a saturated lipid-rich gel phase, a
saturated lipid and cholesterol-rich liquid phase, and an unsaturated
lipid-rich liquid phase. The two liquid phases coexist in some regions of
the phase diagram. The sensitivity of the phase separation to the
components and their composition is demonstrated by the fact that mixtures
of unsaturated lipids and cholesterol, mimicking the components of the
inner leaflet of the plasma membrane, do not phase separate
\citep{wangsilvius}.

Theoretical consideration of mixtures of lipids and cholesterol have been
carried out on binary mixtures of saturated lipids and cholesterol using
models in which the enormous numbers of degrees of freedom of the lipids
are severely reduced. In spite of this simplification, these models have
demonstrated the importance of the preferential interaction between
cholesterol and saturated lipids in bringing about a ``liquid ordered''
phase in which the tails are relatively well-ordered
\citep{ipsen,nielsen}.

It is our contention that two conditions are sufficient for bringing about
coexistence between a cholesterol and saturated lipid-rich liquid ordered
phase and an unsaturated lipid-rich ``disordered liquid'' in which the
tails are typically disordered. The first of these is that the system be
below the main-chain transition of the saturated lipid to its gel phase, a
transition driven by the ordering of the lipid tails. This transition is
first-order, and is characterized by discontinuities in those
thermodynamic quantities which are first derivatives of the free energy
with respect to its arguments; {\em e.g.} the entropy, the area per head
group, etc.  In particular, upon introduction of an unsaturated lipid,
discontinuities arise in the areal densities of the two lipids, first
derivatives of the free energy with respect to the chemical potentials. In
other words, phase separation between a saturated lipid-rich gel phase and
an unsaturated lipid-rich disordered liquid phase is a simple consequence
of the main-chain transition. The difference in component concentrations
in the two phases is expected to be large because the presence of the {\em
cis} double bond causes the unsaturated chains to pack poorly with the
saturated chains that are well-ordered in the gel phase. It also follows
from the first-order nature of the main chain transition that the addition
of cholesterol to the saturated lipid will result in a phase separation
between a saturated lipid-rich gel phase and a liquid phase of lipid and
cholesterol. Whether this liquid phase will coexist with the disordered
liquid phase or simply transform
smoothly to it as the composition of a ternary mixture changes depends on how
different the liquids are. Thus the second sufficient condition for liquid-liquid
phase separation is, as noted above, that the cholesterol favor increased
ordering of the saturated lipid tails, so that the disordered tails of the
unsaturated lipid will also tend to be expelled from the ordered fluid
causing coexistence between it and an unsaturated-rich disordered fluid.

In this paper, we consider a mixture of saturated and unsaturated lipids
within a bilayer and show, as contended above, that below the main chain
transition of the saturated lipid, there is a demixing phase separation.
We do this in a model in which the lipid tails are treated microscopically
according to the rotational isomeric states model of Flory \citep{flory}
which permits each $CH_2$ group to be in one of three configurations; the
lowest energy {\em trans}, or thermally excited gauche plus or 
gauche minus. In the latter, the chain configuration exhibits a kink. Due to
thermal excitation and de-excitation, this kink is transient in contrast to
that in the unsaturated chain which is permanent. To model much of the effect
of the interactions between the lipid segments, we follow early work 
\citep{Ben-Shaul85,Gruen85} and constrain the density of the bilayer
interior to be constant,  equal to a value typical of its oily components. As
shown by these authors, imposition of this packing constraint causes the system
to exhibit a single disordered liquid phase, characterized by a certain average
number of gauche bonds per chain, and a certain area per chain. Larger areas
could be attained by increasing the average number of gauche bonds in the
chains, but this increases the energy and free energy of the system. Smaller
areas could be attained by eliminating even more gauche bonds, but this
decreases the chain entropy to the point that the free energy again increases.
Thus the observed liquid phase occurs at an areal density at which an
incremental decrease in entropy of the chains is just compensated by an
incremental reduction in energy of gauche bonds \citep{Szleifer86}.

The constant density constraint is not adequate to bring about the
main-chain transition to a gel phase, a transition which, we have argued,
is one of the sufficient conditions for raft formation. In order to bring
about {\em both} liquid and gel phases, we not only impose the constant
density constraint, but also include a local orientational interaction
between normals to the $CH_2$ planes of adjacent chains, one which tends
to align them.  We expect that inclusion of this explicit orientational
interaction will capture the aligning tendency of the inter-chain packing.
Indeed we find that this interaction does bring about a second minimum in
the free energy, one characterized by fewer gauche bonds than in the
liquid phase, and which we identify therefore, with the gel phase. In
contrast to the liquid phase, an incremental reduction in entropy of the more ordered
chain conformations is compensated here by an incremental decrease in
orientational energy.
  
We first consider a single component system of saturated tails consisting
of fifteen monomeric segments, modeling those of
dipalmitoylphosphatidylcholine, (DPPC). We find a
first-order main chain transition at a temperature, $T_m$, and calculate
the degree of segment order as a function of position down the chain, and
find good agreement with experiment. There is a concomitant change in area
per head group as the chains lengthen in the gel phase. Next, we apply our
technique to a bilayer of unsaturated lipids.  The unsaturated chains we
consider are seventeen monomeric units long, with a {\emph cis-} double bond
located between monomers eight and nine. This resembles the chain structure of
dioleoylphosphatidylcholine, (DOPC). We find no transition for bilayers composed only of
this unsaturated lipid above the melting point of water. Finally, we
consider a mixture of these saturated and unsaturated chains.  At
temperatures below $T_m$, the system phase separates into a saturated
lipid-rich gel phase and an unsaturated lipid-rich disordered liquid
phase. Although there is little experimental data on this mixture, we find
acceptable agreement with that which exists.

This paper is organized as follows. In the next section, we introduce the
theory for a bilayer slab composed of a mixture of two lipid tails, one
saturated and one unsaturated, and also introduce the interaction we employ for this
system. With these definitions, the partition function for the full
mixture is specified. It is then calculated within the framework of self-consistent field theory in Appendix I.  Central to this theory are two self-consistent equations
which specify two local, effective fields. The first enforces the
constant-density constraint, and the second tends to align the local chain
segment orientation.  This subsection is written for a general system of
two lipid species, one with identical saturated tails, the other with
identical unsaturated tails.
In the following subsection, we introduce the
microscopic model we use to describe the lipid tails.

In section III, the general theory is applied to two specific
lipids, DOPC and DPPC. We first consider two different pure bilayers,
those composed of only one or the other of these lipids.  We present
results pertinent to the main-chain transition in the DPPC bilayer and
some geometric, conformational, and thermodynamic statistics of these
membranes calculated in the context of our model. Of note are the
deuterium order parameters for these lipids since they indicate the acyl
chain's average conformation in the bilayer due to packing with
neighboring lipids. In the final subsection, we present the
phase diagram for the system of mixed lipid bilayers and some
conformational statistics calculated in this model.

\section{THE MODEL AND ITS SELF-CONSISTENT FIELD SOLUTION}

\subsection{Theory}

We consider a bilayer membrane of area $A$ composed of $N_s$ saturated and
$N_u$ unsaturated lipids whose tails occupy a volume $V$, the total
volume of the hydrophobic core of both bilayer leaflets.  Head groups are
placed symmetrically, facing the external, aqueous environment of solvent.
Temporarily, we ignore all interactions between head groups and the solvent
molecules focusing only on the hydrophobic core of the bilayer. Acyl tails
are tethered to the glycerol backbone of the lipid head groups on both
sides of the bilayer, and the first monomer segments extending from the
backbone define the interfaces of the hydrophobic core.  The full width of
this hydrophobic core, including both leaflets, is $L$.  The system of
acyl tails comprising the hydrophobic core is treated as incompressible.
This assumption constrains the average density of monomers to a constant
liquid-like value throughout the bilayer.  The total volume is simply the
sum of all monomer volumes. 

In the following, we limit our attention to lipids consisting of two
identical tails.  We further simplify the system to
that of a mixture of the individual, interacting chains rather than a
system of chain pairs extending from a common headgroup. This should make
little difference to the chain conformations because the chains are so
tightly packed \citep{Fattal94}.

The saturated lipid has the tail structure, $P-(CH_2)_{n_s-1}-CH_3$, which
we model as a chain of $n_s$ linked segments extending from the
glycerol
backbone labeled $P$.  All monomers except the last are $CH_2$ segments
with monomeric volumes $\nu_s(k)=\nu_0=28$\AA$^3,\ k=1,n_s-1.$ The last 
monomer is a $CH_3$ segment,
which is assigned a volume ${\nu_s(n_s)=2\nu_0}$
\citep{Tanford80,Feller98,
Nagle00}. Other than 
these
volumes, the monomers are without structure, and spaced a bond length of
$l_{0}=1.53$\AA \ apart.  The microscopic monomer number density of the
saturated lipid, averaged over the plane of the bilayer, is
\begin{eqnarray}
\label{capphi}
{\hat \Phi}_{s}(z, 
\Sigma)&=&\frac{1}{A}\sum_{\gamma}^{N_s}{\hat \phi}_{s,\gamma}(z,
\alpha_{\gamma}),\\
\label{phi}
{\hat \phi}_{s,\gamma}(z,\alpha_{\gamma})&=&
\sum_{k=1}^{n_s}
{\nu_s}(k)\delta(z-z_{k,\gamma}(\alpha_{\gamma})),
\end{eqnarray}
\noindent where $z_{k,\gamma}(\alpha_{\gamma})$ is the $z$ 
coordinate of the $k$'th monomer
of the ${\gamma}$'th saturated chain when the chain is in the
configuration ${\alpha_\gamma}$.  The circumflex denotes that a quantity
depends upon the lipid configuration. Here the single chain density,
${\hat\phi}_{s,\gamma}$, depends upon the single chain configuration
$\alpha_\gamma$, and the total chain density, $\Phi_s$ depends upon the
configuration $\Sigma$ of all the lipids.
For typographical convenience, we will simply
write ${\hat \Phi}_s(z)$ henceforth, 
suppressing the explicit dependence on
the lipid configurations and allowing the circumflex to remind the
reader of this dependence. We average over the plane of the bilayer as
we shall be considering phases which are translationally invariant in
this plane. 

The unsaturated lipid tail is similarly constructed of $n_u$ linked 
monomers.
It has the tail structure $P-(CH_2)_{x}-CH=CH-(CH_2)_{y}-CH_3$, which is a
monounsaturated chain of $n_u=x+2+y+1$ segments.  Each $CH_2$ and $CH_3$
segment is assigned a volume identical to that in 
the saturated lipid;  
each half of the $-CH=CH-$ segment of the \emph{cis}-unsaturated bond is 
assigned a
volume of $\nu_{CH}={0.8 {\nu_0}}$ \citep{Feller98}.  The microscopic 
monomer number density
of the unsaturated lipid tails is, then, 
\begin{eqnarray} {\hat \Phi}_{u}(z
)&=&\frac{1}{A}\sum_{\zeta}^{N_u}{\hat\phi}_{u,\zeta}(z)\\
{\hat\phi}_{u,\zeta}(z)&=&\sum_{j=1}^{n_u}\nu_u(j)
\delta(z-z_{j,\zeta}),
\end{eqnarray}
\noindent with volume weights,
\begin{eqnarray}
{\nu_u(j)}  = \left\{ \begin{array}{ll}
 {\nu_0} & \textrm{if $j \leq x \quad \textrm{or} \quad x+2 < j < n_u$}\\
 {0.8 {\nu_0}} & \textrm{if $j = x+1, x+2$}\\
 {2 {\nu_0}} & \textrm{if $j = {n_u}.$}\\
  \end{array} \right.
\end{eqnarray}
With these monomer volumes, the total hydrophobic
volumes of single DMPC, DPPC, and DOPC lipids are $784$\AA$^{3}$,
$896$\AA$^{3}$ and $985.6$\AA$^{3}$, respectively, which agree well with the
experimental values of $782$\AA$^{3}$, $913$\AA$^{3}$ and $984$\AA$^{3}$ for the
fluid phases \citep{Nagle00}. We assume these monomeric volumes have no
dependence on temperature.

Much of the effect of the long-range van der Waals attraction and hard
core repulsion is taken into account by the incompressibility constraint.
This is clear from the agreement between results obtained with this
approximation and molecular dynamics simulation for the liquid phases
\citep{Ben-Shaul85,Fattal94}. However an approximation based on density alone cannot
capture the enhanced alignment of the tails which drives the main chain
transition. For this we will need to know the local alignment of the
chains. Our presumption is that the microscopic interactions within
the system favor those configurations in which the tails are more aligned
with one another because in such configurations the chains are more efficiently
packed.  We shall include a pairwise interaction ${\cal E}$ between
chains that reflects
this assumption.

The local orientation of the chains is 
conveniently specified by the normal to the local $CH_2$ group. In 
particular, the normal to the plane determined by the $k$'th $CH_2$ 
group on chain $\gamma$ can be written
\begin{equation}
{\bf u}_{k,\gamma} = 
\frac{{\bf r}_{k-1,\gamma}-{\bf r}_{k+1,\gamma}}{| {\bf r}_{k-1,\gamma}
-{\bf r}_{k+1,\gamma}|}, \qquad k=1...n_s-1. 
\end{equation}
The position vector of the 
anchoring carbon of chain $\gamma$ is denoted ${\bf r}_{0,\gamma}$. 
Subsequently we shall refer to the ${\bf u}_{k,\gamma}$ simply as  
``normals", or ``orientation vectors". 

It is now
convenient to introduce a local density of these normals \citep{schmid} 
which,
for the saturated lipid is,
\begin{equation}
\label{bo}
{\hat \Psi}_{s}(z,{\bf u})=\frac{1}{A}
\sum_{\gamma}^{N_s}\sum_{k=1}^{n_s-1}
\nu_s(k)\delta(z-z_{k,\gamma})
\delta({\bf u}-{\bf u}_{k,\gamma}).
\end{equation}
\noindent 
Again the circumflex is a reminder that this density depends explicitly 
on the chain coordinates
$z_{k,\gamma}$ and normals ${\bf u}_{k,\gamma}$ in a particular
configuration of lipids. 
There is a similar expression, ${\hat\Psi}_u$ for the local density of
normals coming from unsaturated chains.

A local pairwise interaction between tail segments which 
depends upon their orientations can be written, 
{\setlength\arraycolsep{2pt}
\begin{eqnarray}
\lefteqn{{\cal E}[{\hat \Psi_s},{\hat \Psi_u}]  = } \nonumber \\
 && \frac{A}{2 \nu_0}\int dz\int
d{\bf u}d{\bf u'}[{\hat \Psi_s}(z,{\bf u})+
{\hat \Psi_u}(z,{\bf u})]V({\bf u},{\bf u'})
[{\hat \Psi_s}(z,{\bf u'})+{\hat \Psi_u}(z,{\bf u'})].
\end{eqnarray}}
\noindent Here, $\int d{\bf u}$ denotes an integration over the full solid
angle. It is important to note that all interactions are local, \emph{i.e.}
between bonds, and are of the same form and strength irrespective
of the length or degree of unsaturation of the chains to which the 
bonds belong.

Because the normal to the plane of the bilayer, ${\bf c}$, singles out a 
spatial direction, we expect the interaction to depend upon the angles that 
the segments make with this normal
$V({\bf u},{\bf u'})=V({\bf u'},{\bf u})\to V({\bf u}\cdot{\bf c},{\bf
u'}\cdot{\bf c}).$ Irrespective of the particular form chosen for this 
interaction,
within the mean-field approximation which we
introduce shortly, the dependence of the free energy of a single chain 
on the normal  
can only arise through functions of ${\bf
u}\cdot {\bf c}$. It follows from this and the fact that the interaction
energy, ${\cal E}$, is pairwise that the
free energy of the system of chains can only depend upon a product of
such functions. Anticipating this, we take for the interaction a simple form
which already exhibits this product nature,  
\begin{equation}
V({\bf u}\cdot{\bf c},{\bf u'}\cdot{\bf c})
\approx -J f({\bf u}\cdot{\bf c}) f({\bf u'}\cdot{\bf c}) = 
-J f(\cos \theta)f(\cos \theta'),
\end{equation}
where $f(\cos\theta)$ is normalized. Further, in the absence of 
headgroup interactions which we have ignored, 
 the chains will align normal to the plane of 
the bilayer
which implies that $f(\cos\theta)$ be chosen a monotonically decreasing 
function of its argument. With this orientational 
interaction, the lowest energy configuration of two
neighboring orientational vectors 
is one in which they avoid the penalty of hard-core
interaction by both aligning with the bilayer normal.

With this choice of orientational interaction, it is convenient to define 
{\setlength\arraycolsep{2pt}
\begin{eqnarray}
\label{capxi}
{{\hat \Xi}_{s}}( z) & \equiv & \int d\theta \sin \theta 
{\hat \Psi}_{s}(z, {\bf u})f({\bf u}\cdot{\bf c})
 \nonumber\\
&=&\frac{1}{A}\sum_{\gamma=1}^{N_s}{\hat\xi}_{s,\gamma}(z),
\end{eqnarray}
\begin{equation}
\label{xi}
{\hat\xi}_{s,\gamma}(z)
  =  \sum_{k=1}^{n_s-1}{\nu_s}(k)
\delta(z-z_{k,\gamma})f({\bf u}_{k,\gamma}\cdot{\bf c}),
\end{equation}
\noindent with a similar expression for the unsaturated lipid tails, 
so that the interaction energy can be written in the transparent form,
\begin{equation}
\label{interx}
{\cal E}=-{\frac{JA}{2\nu_0}} \int dz [{{\hat \Xi}_{s}(z)}+{{\hat
\Xi}_{u}(z)}]^2.
\end{equation}

For $f(\cos\theta)$ we choose the simple functional form
{\setlength\arraycolsep{2pt}
\begin{eqnarray}
f(\cos \theta) & = & \frac{2m+1}{2}(\cos^{2}\theta)^m \nonumber\\
 & \approx & m\exp(-m{\theta}^2), 
\end{eqnarray}} 
\noindent 
which is approximately a Gaussian with width
${m}^{-\frac{1}{2}}$. The prefactor is a normalization coefficient. 
If the interaction, Eq.  (\ref{interx}), were
not local but were an average over the lipid
chain, and if $m$ were unity in the expression above, then this
interaction 
would be comparable to the
leading order term of the interaction Marcelja \citep{marcelja}
considered for saturated lipids.
The local form, and higher power dependence of the alignment with the bilayer
normal was first considered by Gruen \citep{Gruen85}.

Two parameters determine this interaction between segments, its strength
$J$ and its angular range. The former is set by requiring that the
main-chain transition temperature for the chains under consideration agree
with experiment. The latter may be estimated by considering the average
angular alignment of an acyl tail in the bilayer. Crudely, a lipid in the
bilayer is, on average, confined to a cylindrical shape. A simple estimate of
its alignment with the normal is then $\theta_0\sim {\bar
a}^{\frac{1}{2}}/{\bar l}$, where ${\bar a}$ is an average cross-sectional area
for an acyl chain and ${\bar l}$ is an average length in a typical membrane
leaf. Comparison of  the Gaussian width to this estimate, leads to $m\sim14-18.$ We
use $m=18$ in our calculations.

Now that the interactions between chains are defined, the Helmholtz free energy of
the system, subject to the incompressibility constraint of constant
density, can be obtained within self-consistent, or mean field,  theory. This is done in
Appendix I with the result
\begin{eqnarray}
\label{freeenergy}
\frac{\beta F_{mft}(T,N_s,N_u,A)}{A}&=&-\rho_s\ln Q_s-\rho_u\ln Q_u+
\frac{\beta J}{2\nu_0}\int\ dz 
[\rho_s <{\hat\xi}_s(z)>_s+\rho_u<{\hat\xi}_u(z)>_u]^2
\nonumber\\
&+&\frac{\rho_s}{2}\ln \frac{\rho_s\nu_0}{\rho_s\sum_k\nu_s(k) +\rho_u\sum_k\nu_u(k)}
+\frac{\rho_u}{2}\ln\frac{\rho_u\nu_0}{\rho_s\sum_k\nu_s(k) +\rho_u\sum_k\nu_u(k)}\nonumber\\
&-&\frac{1}{\nu_0}\int\ dz\  \pi(z),
\end{eqnarray}
where $\rho_s\equiv N_s/A$ and $\rho_u\equiv N_u/A$ are the areal
densities of saturated and unsaturated chains.
Here $<{\hat\xi}_s(z)>_s$ is the average in the ensemble of a single 
saturated chain 
\begin{equation}
\label{average}
<{\hat\xi}_s(z)>_s\equiv Tr_{\{\alpha\}}P_s{\hat\xi}_s(z)
\end{equation}
where the trace is over all configurations of a single saturated chain,
and the probability distribution $P_s$ is given by
\begin{equation}
\label{weight}
P_s=\frac{1}{Q_s}\exp\left\{-\beta {\hat
H}_1-\frac{1}{\nu_0}\int[{\hat\phi}_s(z)\pi(z)
+{\hat\xi}_s(z)b(z)]dz\right\}
\end{equation}
with  $Q_s$  the
single saturated chain partition function
\begin{equation}
\label{slpf}
Q_s(\pi,b)= Tr_{\{\alpha\}}\exp\left\{-\beta {\hat
H}_1-\frac{1}{\nu_0}\int[{\hat\phi}_s(z)\pi(z)
+{\hat\xi}_s(z)b(z)]dz\right\},
\end{equation}
and ${\hat H}_1$ the intra-chain part of the Hamiltonian.
The self-consistent equations which determine the fields $\pi(z)$ and
$b(z)$ are
\begin{eqnarray}
\label{scf1}
1&=&\rho_s<{\hat\phi}_s(z)>_s+\rho_u<{\hat\phi}_u(z)>_u,\\
\label{scf2}
b(z)&=&-\beta J[\rho_s<{\hat\xi}_s(z)>_s+\rho_u<{\hat\xi}_u(z)>_u].
\end{eqnarray}
The first of these enforces the incompressibility constraint.

Note that, because of the incompressibility constraint, the volume is
not an independent variable so that the Helmholtz free energy,
Eq. (\ref{freeenergy}),
depends only on the temperature, the numbers of each lipid, and the area.  
The interpretation of the free energy is straightforward.
The first two terms are simply the single chain Gibbs free
energies multiplied by the number of chains per unit area. 
As is well known, these terms  double-count the pairwise interaction energy,
and this is corrected by the third term. The next two terms are the entropy
of mixing, and the last is simply $-\beta pV/A$ which relates the Gibbs
free energy of the preceding terms to a Helmholtz free energy. 

Other than excluding tail configurations protruding into the solvent
region, the effects of the aqueous medium and head groups have thus far
been neglected. It is evident that the interfacial effects can also be
somewhat complex. For example, tilted solid phases are often attributed to
a mismatch between a smaller average areal footprint of the packed tails
and a larger headgroup. On the other hand, a small headgroup mismatched
with larger, fluid-like tails attached to it can yield costly interfacial
interactions between the solvent and the hydrophobic core, resulting in a
large interfacial tension. These effects are beyond the scope of our
calculation. However, so as not to overlook the contributions from the
head groups completely, we choose a minimal model for them.

Ignoring the molecular details of the head groups and their interactions
with the aqueous environment, we focus on two physical attributes that
result from these interactions. The first is an interfacial energy per
unit area, which we include as a simple phenomenological form,
${\gamma_0}A$. Thus the total free energy per unit area is
\begin{equation}
\label{finalfreeenergy}
f_A(T,\rho_s,\rho_u)=F_{mft}/A+\gamma_0,
\end{equation}
where $F_{mft}$ is given in Eq. (\ref{freeenergy}). 
The value of $\gamma_{0}$ is taken to be on the order
of the energy per unit area of an oil-water interface. We use
$\gamma_{0}=0.12k_{B}T/$\AA$^2$ in our calculations, which is roughly the
experimental value. The second effect attributable to the head groups 
is the rigidity they provide the acyl
tail. The lipid tails are tethered to the glycerol backbone, providing a fixed
endpoint for the chain. The head groups tend to order the first segments
of the lipid tethered there, an
effect which we model by a surface field that couples to the first bond of the
lipid. To model this, we include in ${\hat H}_1$ a surface field 
term $\eta \cos({\bf
u}_1\cdot{\bf c})$.  The 
magnitude of the surface field $\eta$ is tuned so that the average 
angle of the first bond is
approximately equal to that obtained from NMR measurements for the lipid.
 
\subsection{Microscopic Model}

To evaluate the single lipid partition function in external fields, Eq.
(\ref{slpf}) above, we specify a microscopic model suitable for describing the
noninteracting single chain configurations and their energy, ${\hat H}_1$. 
To this end, we employ Flory's
Rotational Isomeric States (RIS) model \citep{flory} for the lipid acyl chains. This
model places a subsequent monomer in the acyl chain at one of three discrete
possible orientations from its ancestor segment; the
trans, gauche plus, and gauche minus orientations. According to
this model, the gauche bonds are local excitations that contort the chain and
increase the energy by $\epsilon_g = 500$ cal/mol. Thus ${\hat H}_1$ of a given
configuration is simply equal to $\epsilon_g$ multiplied by the number of
gauche bonds in that configuration, and also includes the contribution from the
surface field described above. The lipid conformation with all trans bonds is a
highly stretched chain, whereas a lipid tail with several
gauche bonds is coiled. Given the location of the first bond in the acyl chain,
all positions of the subsequent $n_s-1$ generations are determined by this
model, which locates segments in a tetrahedral fashion, with skeletal angles of
$112^{\circ}$, and dihedral angles of $\{0^{\circ},\pm120^{\circ}\}$ for trans
and gauche$\pm$ orientations respectively. Without restrictions on this set,
there are $3^{n_s-1}$ configurations for each fixed, first bond orientation.

The \emph{cis-} double bond in the unsaturated chains
slightly modifies this model.  This bond is a quenched defect in the
chain, effectively placing a kink in the tail about the location of the
double bond.  The energies and configurations of the segments adjoining
the double bond as well as the $-C=C-$ bond length differ slightly from
the saturated links. For the segments associated with the double bond, we
employ the RIS configurational model of \emph{cis}-1,4-polybutadiene
\citep{mark}, which possesses the same atomic structure as the double bonds in
these lipid chains.

We also impose several restrictions on the lipid configurations that reduce the
number of allowed ones significantly.  First of all, configurations that
protrude into the aqueous environment are discarded. A configuration with any
two, non-consecutive monomer centers less than a bond width apart is considered
self-intersecting and also discarded. Configurations with consecutive
gauche minus (plus), gauche plus (minus) bond sequences have large
hydrogen-hydrogen steric overlaps \citep{flory}, a phenomena referred to as the
``pentane effect.'' These configurations are also neglected. Each single lipid
partition function is then evaluated by enumerating all allowed configurations
in the RIS model, from many different first bond orientations and locations. It
should be noted that two different first segment orientations will have different
numbers of configurations stemming from them since they will have different
number of configurations that penetrate the aqueous environment. 

We specify in Appendix II how the probability distributions $P_{s}$ and
$P_u$ and the single lipid partition functions, Eq. (\ref{slpf}), are
calculated within the context of the RIS model and the configurational
restrictions. We also discuss there a few other details of the calculation of
the free energy. 

\section {RESULTS} 
\subsection{Single component DPPC and DOPC bilayers}

We first apply our method to a tensionless bilayer composed only 
of DPPC lipids. We
solve the self-consistent equations for a given temperature and under the
condition that the surface tension be
zero. The high temperature liquid phase
undergoes a main chain transition to a low temperature gel phase, and  by
adjusting the interaction strength $J$, we set this transition temperature at
$T_{m}\sim 315$\ K, essentially the experimental one \citep{finegold}.
The value of $J$ remains fixed at this value throughout all subsequent 
calculations. In Fig. 1 we show the Helmholtz free energy {\em per
two-chain lipid}  
\begin{equation}
f_{N_s}(T,a_s)=2\left(\frac{F_{mft}}{N_s}+\gamma_0a_s\right)
\end{equation}
in units of $k_BT$ plotted {\em vs.} the area per two-chain lipid, 
$2A/N_s=2a_s$, at the
temperature of the main chain transition.
It is straight forward to show that
\begin{eqnarray}
\label{fperlipid}
f_{N_s}(T,a_s)&=&2(\mu_s+\gamma a_s)\\
\label{dfperlipid}
df_{N_s}&=&2[\gamma da_s-(S/N_s)dT],
\end{eqnarray}
with $S$ the total entropy.
The two minima in the figure occur at the areal densities of the two
coexisting phases. Because these states are minima of the free energy
per lipid, their surface tensions vanish, as seen from
Eq. (\ref{dfperlipid}). Further as they have the same free energy per
lipid at zero surface tension, they have the same chemical potential, as
seen from Eq. (\ref{fperlipid}). Thus these phases are indeed at coexistence.
The main chain transition is clearly seen to be first order, marked by
a discontinuous jump in
the area per lipid.  Table \ref{table:res} summarizes a few
geometric aspects of each state and of the transition. We find satisfactory
agreement with experiment for the geometric properties.  The fluid phase is
characterized by a greater area per lipid than is the gel phase. The
hydrophobic width of the bilayer is thinner, filled with coiled lipid tails
with, on average, somewhat more than four gauche bonds per tail. The gel state
consists of more ordered lipid chains that are tightly packed, having a smaller
area per lipid and fewer gauche bonds. 

One thermodynamic quantity that is sensitive to the packing in the hydrophobic
core is the deuterium ``$S_{CD}$'' order parameter profile, which is commonly
measured in the fluid state by nuclear magnetic resonance (NMR) experiments.
The calculated deuterium order parameter for our model DPPC bilayer is shown in
Fig. 2 for both fluid and gel states. The fluid phase is  
compared to experimental data from reference 20. Collective
organization of the lipid tails during the transition from the fluid to gel
state is indicated by the large jump in chain alignment. Experiment and
simulation distinguish $sn-1$ and $sn-2$ lipid tails, which we do not, so the
calculated profiles in Fig. 2 should be considered as an
average over both acyl tails.  Comparing to experiment, we see that the
calculated fluid phase is slightly more disordered than the experimental one.

Applying our calculational method to a bilayer consisting only of
mono-unsaturated, DOPC model lipids, we find a fluid phase and no transition
above the freezing point of water. This result is consistent with experiment,
which indicates a transition temperature for DOPC of $T_m\sim250$ K. The
calculated fluid state has an area per lipid of $70.4$\AA$^2$ and a hydrophobic
width of $28.0$\AA. This lipid area is somewhat smaller than the
experimental value of $72.5$\AA$^2$ with also a thinner hydrophobic width of
$27.1$\AA.  The calculated deuterium order parameter is shown in Fig.
2. It should be noted that these profiles are calculated for a
lower temperature, $T=305$ K, than the saturated lipids.  The order parameter
displays a notch in the profile about the \emph{cis-} double bond, one which is
also seen in MD simulation \citep{fellerDOPC,scottDOPC1,scottDOPC2}. Also,
there is a slight even-odd effect in the order parameter which we do not
observe in the saturated chains. 

\subsection{Composite bilayers of DPPC and DOPC}

From the calculation of the free energy, Eq. (\ref{freeenergy}), the phase
behavior for a mixture of both lipids, DPPC and DOPC, can be determined. Phase
coexistence is determined by examining the auxiliary function,
\begin{equation}
{\tilde\gamma}= 
\min_{{\rho_s},{\rho_u}}
\left\{ { f_{A}-{\lambda}_{s}{\rho}_{s}-{\lambda}_{u}{\rho}_{u}}\right\}.
\end{equation}
\noindent When the $\lambda_s$ and $\lambda_u$ are adjusted so that
$\tilde\gamma$ exhibits two minima with the same value of $\tilde\gamma$, then there
is two-phase coexistence.  At this point, the values of $\lambda_s$,
$\lambda_u$ are  equal to the values of the chemical potentials of the
saturated and of the unsaturated lipids, $\mu_s$ and $\mu_u$, in the coexisting
phases and $\tilde\gamma$ is equal to the surface tension, $\gamma$, of each
phase.  We further require that this surface tension be zero in each phase as
there is no constraint on the bilayer area. The four requirements that the
chemical potentials and surface tension be equal in each phase and that the
tension be zero determines the two areal densities, $\rho_s$ and $\rho_u$, in
each of the two phases.  Note that the bare surface tension, $\gamma_0$, causes
the system to contract to the point at which the tension due to packing just
counteracts the bare tension leaving a net tension of zero.

The calculated phase diagram of the system is shown in Fig.
3.  For small concentrations of DOPC, we find a gel state
rich in DPPC below the main-chain transition temperature. A region of fluid/gel
coexistence is  found which  is rather wide because of the poor packing of the
unsaturated lipids. These results are in reasonable agreement with 
experiment \citep{lentz}. 

The deuterium order parameters for the two coexisting phases at $T=293$ K
are shown in Fig. 4. The gel phase shows significant ordering
of both lipids, given by the large absolute magnitude of both order parameters.
The disordering effect of the double bond is more prominent in the ordered
phase, with affected bonds showing a large notch. This is noteworthy because
it implies that the tight packing about the unsaturated lipids in the ordered phase tends to
\emph{enhance} the disorder about the unsaturated bond.  The
order parameters in the liquid-crystalline phases are also shown 
in Fig. 4.

\section {SUMMARY}
We have formulated a microscopic model of a bilayer consisting of lipids, one
which incorporates two important effects of the interactions between chains;
the tendency of the chains to create an incompressible hydrophobic interior,
and their cooperative tendency to order. The solution of the model  obtained
from self-consistent field theory reproduces many salient features of the lipid
bilayer.  The system of chains which mimic DPPC undergoes a first-order main
chain transition from a disordered fluid state to a more ordered gel phase, a
transition characterized by a reduction in the number of gauche bonds and the
area per chain.   The
ordering of the chains in the liquid phase as determined by the deuterium order
parameter is in rather good agreement with experiment. We have also
calculated the deuterium order parameter in the gel phase. 

We then applied the model to a system mimicking the mono-unsaturated chains of
DOPC.  Utilizing the same interaction as before, we find that this system does
not undergo a main chain transition above zero degrees centigrade, a result in
agreement with experiment.  The lowering of the main-chain transition is  due
to the double bond in the chain which disrupts the tendency to order. The
deuterium order parameter profiles for this system show the characteristic
reduction at the position of the double bond.

Lastly we examined a mixture of saturated and unsaturated lipids and showed
that the system phase separates into a saturated-lipid-rich gel phase and an
unsaturated-lipid-rich liquid phase at temperatures below the main-chain
transition temperature of the saturated lipid. The separation arises
from the first-order main chain transition. The large difference in
areal densities is due
to the lipid architecture; the unsaturated lipids are rejected from the gel
phase because they do not pack into it very well.
 
 Several pieces of physics have been ignored in our calculation, such as the
 explicit interaction between head groups and, of course, the effect of 
 fluctuations. We
 expect that such effects will alter the specific values of thermodynamic
 quantities, but not the main phenomena. In particular the phase separation
 into a saturated-rich gel and an unsaturated -rich liquid phase will remain
 below the main chain transition of the saturated lipid.  As noted earlier,
 this is important as this behavior is one of two ingredients for a sufficient
 theory of liquid-liquid coexistence in a ternary mixture which includes
 cholesterol. We are currently investigating the role of this third component
 within the framework of our model.
 
 \section {ACKNOWLEDGMENTS}

We thank Sarah Veatch, Sarah Keller, Ben Stottrup, Kazuya Okubo, and
Marcus M{\"u}ller for
stimulating conversations, and Avinoam Ben-Shaul for helpful correspondence.

This material is based upon work supported by the National Science
Foundation under Grants No. 140500 and CTS-0338377, and in 
part by a National Science Foundation IGERT fellowship
from the University of Washington Center of Nanotechnology.

\section {APPENDIX I. SELF-CONSISTENT FIELD THEORY} 
The exact Helmholtz free energy of the system is
\begin{equation}
\label{exactf}
F=Tr_{\{\Sigma\}}P_{ex}\left[H+\beta^{-1}\ln\ P_{ex}\right]
\end{equation}
where the trace is a sum over all distinguishable 
configurations of all chains, two chains per molecule, and
$P_{ex}$ is the exact probability distribution function for the
system. It satisfies
\begin{equation}
\label{exact}
 Tr_{\{\Sigma\}}P_{ex}=1.
\end{equation}
The Hamiltonian of the system is
\begin{equation}
H=\sum_{\gamma=1}^{N_s}{\hat H}_{1,\gamma}+\sum_{\eta=1}^{N_u}
{\hat H}_{1,\eta}
-\frac{JA}{2\nu_0}\int\ d{z}[\frac{1}{A}\sum_{\gamma=1}^{N_s}\hat\xi_{s,\gamma}
(z)+\frac{1}{A}\sum_{\eta=1}^{N_u}\hat\xi_{u,\eta}(z)]^2,
\end{equation}
where ${\hat H}_{1,\gamma}$ contains the intra-chain contributions to the
energy of chain $\gamma$, such as those arising from gauche bonds 
and the coupling to the surface field.
Self-consistent, or mean field, theory consists of approximating $P_{ex}$ by
the product of one chain probability distributions,
\begin{equation}
\label{approx}
P_{ex}\approx \left(\frac{N_s}{2}\right)!\left(\frac{N_u}{2}\right)!\prod_{\gamma=1}^{N_s}\prod_{\eta=1}^{N_u}
P_{s,\gamma}
P_{u,\eta}
\end{equation}
where the probability distribution of the single saturated chain is
written   
\begin{eqnarray}
\label{probdist}
P_{s,\gamma}&=&\frac{1}{Q_s}\exp\left\{-\beta {\hat
H}_{1,\gamma}-\sum_k\frac{\nu_s(k)}{\nu_0}\left[\Pi(z_{k,\gamma})+B_s(
z_{k,\gamma})f({\bf u}_{k,\gamma}\cdot{\bf c})\right]\right\},\\
&=&\frac{1}{Q_s}\exp\left\{-\beta {\hat
H}_1-\frac{1}{\nu_0}\int[{\hat\phi}_s(z)\Pi(
z)+{\hat\xi}_s(z)B_s(z)]dz\right\}.
\end{eqnarray}
with the fields $\Pi$ and $B_s$  to be determined, and $Q_s$  the
single saturated chain partition function
\begin{eqnarray}
Q_s(\Pi,B_s)&=&Tr_{\{\alpha\}}\exp\left\{-\beta {\hat
H}_1-\sum_k\frac{\nu_s(k)}{\nu_0}\left[\Pi(z_{k})+B_s(
z_{k})f({\bf u}_{k}\cdot{\bf c})\right]\right\},\\
            &=& Tr_{\{\alpha\}}\exp\left\{-\beta {\hat
H}_1-\frac{1}{\nu_0}\int[{\hat\phi}_s(z)\Pi(
z)+{\hat\xi}_s(z)B_s(z)]dz\right\}.
\end{eqnarray}
Here 
\begin{equation}
{\hat\phi}_s(z)=\sum_{k=1}^{n_s}\nu_s(k)\delta(z-z_k),
\end{equation} 
defined previously in  Eq. (\ref{phi}), 
the trace is over all configurations of a single saturated 
chain, and the 
index specifying the particular chain has been dropped. The expressions
for $P_u$ and $Q_u$ are essentially the same except that the segment volumes
are those of the unsaturated chain, and the trace is over its configurations.
Note that because
\begin{equation}
Tr_{\{\alpha\}}P_s=1,
\end{equation} 
and similarly for $P_u$, and because the sum in the normalization of the
exact probability distribution, Eq. (\ref{exact}), is
over all {\em distinct} states,
the factor of $(N_s/2)!(N_u/2)!$ is needed in Eq. (\ref{approx}) so that the
normalization 
will be satisfied within the mean field approximation.

The Helmholtz free energy of Eq. (\ref{exactf}) can now be obtained directly within
the mean field approximation of Eq. (\ref{approx}). One obtains
\begin{eqnarray}
\frac{\beta F}{A}&=&-\frac{1}{\nu_0}\int\{\frac{\beta J}{2}[\rho_s<{\hat
\xi}_s>_s+\rho_u<{\hat\xi}_u>_u]^2+\rho_s<{\hat\xi}_s>_sB_s(z)
+\rho_u<{\hat\xi}_u>_uB_u(z)\nonumber\\
&+&[\rho_s<{\hat\phi}_s>_s+\rho_u<{\hat\phi_u}>_u]\Pi(z)\}dz\nonumber\\
&-&[\rho_s\ln\ Q_s+\rho_u\ln\ Q_u]
+\frac{\rho_s}{2}\ln\ N_s+\frac{\rho_u}{2}\ln\ N_u,
\end{eqnarray}
where $\rho_s\equiv N_s/A$ and $\rho_u\equiv N_u/A$ are the areal
densities of saturated and unsaturated chains,
\begin{equation}
<{\hat\phi}_s>_s\equiv Tr_{\{\alpha\}}P_s{\hat\phi}_s,
\end{equation}
and similarly for the other single-chain ensemble averages. Again chain
indices have been dropped.

The unknown fields $\Pi(z)$, $B_s(z)$, and $B_u(z)$ are
obtained from three conditions. The first is that the interior of the
bilayer be incompressible which implies
\begin{equation}
\label{inc}
\rho_s<{\hat\phi}_s(z)>_s+\rho_u<{\hat\phi}_u(z)>_u=1.
\end{equation}
The second and third are that the free energy be an extremum with respect
to the undetermined functions $<{\hat\xi}_s>_s$ and
$<{\hat\xi}_u>_u$. This leads to the equations
\begin{eqnarray}
B_s(z)&=&B_u(z)\\
\label{beq}
            &=&\beta J[\rho_s <{\hat\xi}_s(z)>_s+\rho_u<{\hat\xi}_u(z)>_u].
\end{eqnarray}
We shall denote the functions which satisfy the self-consistent 
Eqs. (\ref{inc}) to
(\ref{beq}) as
$\pi(z)$ and $b(z)$ respectively.

Substituting the fields $\pi(z)$ and $b(z)$ into the free energy, one obtains
\begin{eqnarray}
\frac{\beta F}{A}&=&-\rho_s\ln Q_s-\rho_u\ln Q_u+\frac{\beta J}{2\nu_0}\int\ dz 
[\rho_s <{\hat\xi}_s(z)>_s+\rho_u<{\hat\xi}_u(z)>_u]^2\nonumber\\
&+& \frac{\rho_s}{2}\ln N_s+\frac{\rho_u}{2}\ln N_u -\frac{1}{\nu_0}\int\ dz\ \pi(z).
\end{eqnarray}
The first two terms are simply the single chain Gibbs free
energies multiplied by the number of chains per unit area. (They are
Gibbs free energies as the pressure-like $\pi$ is the independent variable.) 
These terms  double-count the pairwise interaction energy,
and this is corrected by the third term.  The fourth and fifth terms are
the entropy of mixing. Thus the first five terms are the mean field
approximation for the Gibbs free energy per unit area of the system. 
The last term is simply $-\beta pV/A$, with
$p$ the pressure, and relates the Gibbs free energy to the Helmholtz
free energy.  

The contribution of the entropy of mixing can be 
put in a more convenient form as follows. We note 
\begin{equation}
(N_s\sum_k^{n_s}\nu(k)+N_u\sum_k^{n_u}\nu(k))/V=1,
\end{equation}
so that
\begin{equation}
\ln\left[\frac{N_s\sum_k^{n_s}\nu(k)+N_u\sum_k^{n_u}\nu(k))}{\nu_0}\right]=
\ln\left[\frac{V}{\nu_0}\right],
\end{equation}
where $\ln(V/\nu_0)$ is constant within the canonical ensemble. 
We multiply this constant by
$(N_s+N_u)/2$ and subtract it from the free energy above  because this
term, 
linear in $N_s$ and $N_u$,  only shifts the definitions of the chemical
potentials. 
We therefore arrive at the final expression for the mean field free energy
\begin{eqnarray}
\label{mffreeenergy}
\frac{\beta F_{mft}}{A}&=&-\rho_s\ln Q_s-\rho_u\ln Q_u+\frac{\beta J}{2\nu_0}\int\ dz 
[\rho_s <{\hat\xi}_s(z)>_s+\rho_u<{\hat\xi}_u(z)>_u]^2
\nonumber\\
&+&\frac{\rho_s}{2}\ln \frac{\rho_s\nu_0}{\rho_s\sum_k\nu_s(k)
+{\rho_u}\sum_k\nu_u(k)}
+\frac{\rho_u}{2}\ln\frac{\rho_u\nu_0}{\rho_s\sum_k\nu_s(k) +\rho_u\sum_k\nu_u(k)}\nonumber\\
&-&\frac{1}{\nu_0}\int\ dz\  \pi(z).
\end{eqnarray}

\section {APPENDIX II. THE RIS DISTRIBUTION}
This section clarifies how the sums over single lipid configurations, which are
needed to evaluate single lipid ensemble averages, Eq. (\ref{average}), are
carried out within the rotational isomeric states (RIS) model, and how the
calculation of thermodynamic quantities is accomplished.  In the RIS model
\cite{flory}, the bond between the carbon at position ${\bf r}_{k-1}$ and ${\bf
r}_k$ , $k\geq 2$, can be in one of three configurations, the lowest energy
trans  or the  gauche plus or gauche minus with an energy higher
than that of the trans by an amount $\epsilon_g$. There can be no 
gauche bond between the anchoring carbon and the first carbon in the chain by
definition. Further an apparent gauche bond between the first and second
carbons can actually be obtained by a solid body rotation of a  trans
configuration.  Within the RIS model, the sequence of bond orientations down
the chain, the location of the anchoring carbon, ${\bf z}_0$, and the
orientation of the first normal, ${\bf u}_1$, completely define the
configuration of the chain. 

The probability distribution, Eq. (\ref{probdist}), 
of a saturated chain in configuration
$\alpha$ can now be written
\begin{eqnarray}
P_s(\alpha)&=&\frac{1}{Q_s}\prod_{k=1}^{n_s}P_{s,k}(\alpha)\nonumber\\
Q_s&=&
Tr_{\{\alpha\}}\prod_{k=1}^{n_s} P_{s,k}(\alpha)\nonumber  
\\
P_{s,k}(\alpha)&=&\exp(-\beta {\hat H}_{eff,k}),
\end{eqnarray}
where the effective Hamiltonian is defined as
\begin{eqnarray}
\beta {\hat H}_{eff,1}&\equiv& \eta\cos({\bf u}_1\cdot{\bf
c})+\frac{\nu_s(1)}{\nu_0}[\pi(z_1)+b(z_1)f({\bf
u}_1\cdot{\bf c})],\nonumber\\
\beta{\hat H}_{eff,2}&=&\frac{\nu_s(2)}{\nu_0}[\pi(z_2)+b(z_2)f({\bf
u}_2\cdot{\bf c})],\nonumber\\
\beta{\hat H}_{eff,k}&=&g_k\epsilon_g+
\frac{\nu_s(k)}{\nu_0}[\pi(z_k)+b(z_k)f({\bf
u}_k\cdot{\bf c})],\qquad  2<k\leq n_s-1\nonumber\\
\beta{\hat H}_{eff,n_s}
&=&g_{n_s}\epsilon_g+\frac{\nu(n_s)}{\nu_0}\pi(z_{n_s}),
\end{eqnarray}
where  $\eta$ is the  surface field introduced earlier which couples to the
first normal, and $g_k$ equals unity if the  bond between the carbon at ${\bf
r}_{k-1}$ and ${\bf r}_k$ is gauche, and zero otherwise.  For unsaturated
lipids, the \emph{cis}-unsaturated bond alters the bond weights.  These
energetic weights and configurational properties are specified in Mark
\citep{mark}.

Not all possible bond configurations in a chain are acceptable. Those in which
a gauche minus immediately follows a gauche plus or {\em vice versa}
have large  steric overlaps \citep{flory}, and are discarded. Further those
configurations in which the chain intersects itself are also discarded as are
those in which the chain pierces the aqueous  plane and enters the region of
solvent.

To carry out the sum over single lipid configurations, we first enumerate all
configurations of the chains within the RIS model for a {\em given} position of
the first monomer and orientation of the first bond. Under this constraint, the
RIS configurations of these lipids are generated once and stored in a linked
list with a \emph{tree} structure.  This structure allows us to store a large
number of conformations and perform configurational sums using roughly a
quarter of the number of operations required by a linear storage array.  We
then sample orientations and translations of this first bond by rotating it
about the bilayer normal and then translating the chain along the bilayer
normal.  In particular we consider rotations of the lipid configurations as a
solid body, sampling two Euler angles: that to the aqueous plane from the
normal, $\theta$, and $\phi$, and about the body axis. 

Translations of the whole chain normal to the bilayer are chosen in a
small interval ${\{-\Delta z,\Delta z\}}$, with $\Delta z \sim$ 1.5\AA, about
the aqueous plane. We choose the first segments of the chains to originate at
one of four evenly spaced locations in this window about the aqueous plane,
matching a discretization of the bilayer we employ, described below. The
allowed Euler rotations about the first bond varies with these origins. For
example, the maximally inserted chains in this translational grid have somewhat
more than the hemisphere of allowed rotations without penetrating the aqueous
environment, whereas the maximally extracted ones have less than the
hemisphere. 

For each of these origins in the translational window, we sample the allowed
solid angles. Using a shorter saturated lipid, 12 or 14 segments long, as a
benchmark, we may sample both Euler angles exhaustively. A large uniform
sampling, $(\sim \textrm{3500 total angles})$, indicates the dense phase has a
preference for two orientations, $\theta\sim0$, corresponding to
the first bond in the \emph{trans} state, and to a lesser extent,
$\theta\sim60^{\circ}$ corresponding to a first bond gauche
isomerization.  Since such an exhaustive sampling of angles for longer lipids
is not computationally feasible, we must substantially reduce the number of
angles sampled.

One way of accomplishing this is to utilize a set of angles determined from
equidistant points on a sphere. This accomplishes a uniform density
distribution on the sphere, each angle subtending nearly the same portion of
the total solid angle. These angles are determined from a simulation of
repulsive points confined to a sphere. We then increase the density of
configurations for small $\theta$, until we nearly reproduce the statistics
from the exhaustive sampling above. In order to correct for the bias imposed by
the selection of angles, the evaluation of the trace such as in Eq.
(\ref{average}) includes a factor proportional to the amount of the total
allowable solid angle they subtend. The dense set of angles near $\theta\sim0$
each
subtend a smaller amount of area and have a smaller weight.

With these rotations and translations, and discarding self-intersecting and
other restricted configurations, the complete computational set includes on the
order of $10^7$ configurations for roughly 150 total initial bond angles for
each lipid species. 

To determine ensemble averages of position dependent quantities, such as
\begin{equation}
\label{phiapp}
{\hat \phi}_{s,\gamma}(z,z_{k,\gamma})=
\sum_{k=1}^{n_s}
{\nu_s}(k)\delta(z-z_{k,\gamma}),
 \end{equation}
 of Eq. (\ref{phi})
 or  
 \begin{equation}
\label{xiapp}
 {\hat\xi}_{s,\gamma}(z)
  =  \sum_{k=1}^{n_s-1}{\nu_s}(k)
\delta(z-z_{k,\gamma})f({\bf u}_{k,\gamma}\cdot{\bf c}),
\end{equation}
 of Eq. (\ref{xi}),
we discretize the space of the continuous variable $z$ into 2${\cal N}_z$
levels $z_l$ of width $\delta z=L/2{\cal N}_z$. A saturated chain which is
without gauche defects and which is aligned with the bilayer normal sets a
limiting, maximal half-width, $L/2$, for the hydrophobic core of a
one-component lipid bilayer.  Typically, we divide the hydrophobic width into
slices $\delta z\sim0.75$\AA\  to achieve a suitable resolution for our
calculations. Free energy calculations for various bilayer widths are carried
out by changing the number of slices, $2{\cal N}_z$, but maintaining the slice
width $\delta z$. By changing the bilayer thickness incompressibly, we simply
vary the average areal density of the lipids. The equilibrium states
of the system are typically not among the discrete points we calculate, which
may be seen in Fig. (1). Structural properties, such as the $S_{CD}$ order
parameters, are sensitive to the areal density. To estimate the order
parameters, shown in Figs. (2) and (4), we first estimate the areal density of
the phase from the free energy calculation, and then vary $\delta z$ so that
the bilayer acquires nearly the equilibrium density of the phase. In practice,
changing $\delta z$ slightly affects the packing of the chains in the
hydrophobic core so several values of $\delta z$ are used, and the order
parameters are averaged over these various discretizations. This practice
introduces an uncertainty in these order parameters which we estimate as 
$\pm 0.02$. 

The calculation is now straightforward. For given surface compositions $\rho_s$
and $\rho_u$ and temperature $T$,  the densities $<{\hat\phi}_s(z_l)>_s$,
$<{\hat\phi}_u(z_l)>_u$, $ <{\hat\xi}_s(z_l)>_s$, and $ <{\hat\xi}_s(z_l)>_s$
are evaluated in the ensemble with the probability distribution $P_s$ of
Eq. (\ref{weight}) or $P_u$ containing given fields $\pi(z_m)$ and
$b(z_m)$. These fields
must satisfy the self-consistent equations, Eqs. (\ref{scf1}) and
(\ref{scf2}), which contain the densities.  This set of coupled, nonlinear,
equations can be solved using a standard algorithm. Due to the mirror symmetry
of the bilayer,  it is sufficient to solve the self-consistent equations in the
half-bilayer. To do this, chain interdigitation must be correctly accounted for
in lipid configurations that extend past the midplane into the other leaf.
If the configuration is long enough to interdigitate the other leaf, its
monomer densities are reflected and the `image' of the configuration counted in
the statistics.  The free energy, Eq. (\ref{freeenergy}), is then calculated
from which all thermodynamic information may be derived.

The parameters we have chosen for the calculations below are as follows.  The
saturated lipid chains we consider are $n_{s}=15$ monomers long, mimicking
DPPC. An interaction of strength $J/T_m=0.09$ induces the
main-chain transition at its observed value of $T_m=315$ K.  The
unsaturated lipids have $n_{u}=17$ monomers with a \emph{cis-9} double bond,
between the eighth and ninth segments. This is the chain structure of
DOPC.  For both species of lipid, we find a surface field of 
$\eta/{k_B T_m}\sim1.03$ sufficient to obtain reasonable values of the order
parameter for bonds nearest the aqueous plane. 

\clearpage
\bibliography{lipids}

\clearpage

\clearpage

\begin{table}[!t]
\centering
\caption{Mean-Field and Experimental
Results for the DPPC Bilayer. Experimental results are from 
Nagle and Tristam-Nagle\citep{Nagle00}.} 
\begin{tabular*}{6in}{l@{\extracolsep{\fill}}cr}
\hline\hline
Quantity & Exp't & Mean Field \\ [0.5ex]
\hline
Fluid Area/lipid $T=323$ K (\AA$^2$) & 64.0 & 67.0 \\
Gel Area/lipid $T=293$ K (\AA$^2$) & 47.9 & 49.9 \\	
Fluid Hydrophobic width $T=323$ K (\AA) & 28.5 & 26.7 \\ 
Gel Hydrophobic width $T=293$ K (\AA) & 34.4 & 35.9 \\ 
Gauche Bonds (gel, $T=315$ K) & & 2.1  \\ 
Gauche Bonds (fluid, $T=315$ K) & & 4.3 \\ [1ex] 
\hline\hline
\end{tabular*}
\label{table:res}
\end{table}

\clearpage
\section{Figure Captions}

Fig. 1 \
Free energy per two-chain saturated lipid, in units of $k_BT$, as a
function of area per two-chain lipid. The temperature is essentially that of
the main chain transition, approximately $315$ K, at which the two
phases are in coexistence. In each phase the surface tension, $\gamma$,
vanishes. The solid line is to guide the eye.

Fig. 2 \
Various calculated deuterium order parameters for a DPPC and a DOPC
bilayer are compared with experiment and simulation. The top figure shows the
calculated order parameters in coexisting liquid and gel phases for a DPPC
bilayer at the transition temperature. The middle figure compares the
mean-field deuterium order parameter for the fluid phase at ($T=323$ K) to that
obtained from experiment \citep{brown} ($T=323$ K) and from MD simulation
\citep{scottDPPC} ($T=325$ K, sn-2 chains) for the fluid phase. The bottom
figure compares our results for the fluid DOPC bilayer to simulation
\citep{scottDOPC2} at $T_m=305$ K, sn-2 chains. The indices have been shifted
from the text in accord with most experiments and simulations.

Fig. 3\
Calculated phase diagram for a bilayer composed of DPPC and
DOPC as a function of the volume fraction of DOPC. The error bars are an
estimate of the uncertainty in the calculation introduced by the bilayer
discretization. 

Fig. 4\
Deuterium order parameters for a composite bilayer of DPPC and DOPC at
$T_m=300$ K  in each of the coexisting phases. The order parameters for DOPC are
identifiable by the notch half way down the chain. The upper part of the figure
shows the gel phase order parameters, and the lower part shows the liquid phase
order parameters. The indices have been shifted from the text in accord with
most MD simulations  

\clearpage

\clearpage

\begin{figure}[!t]
\includegraphics[scale=0.6]{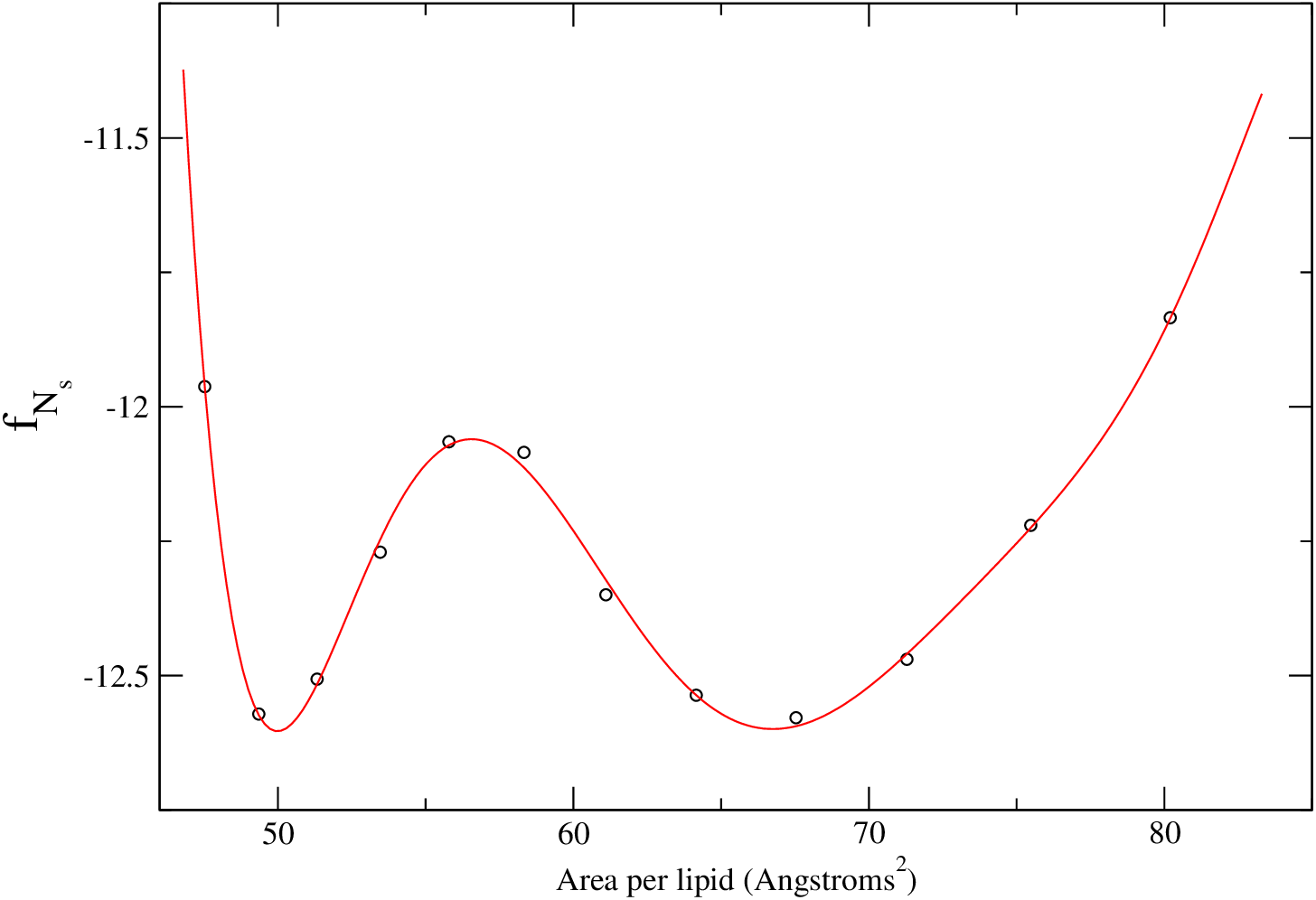}
\label{coexistence}
\end{figure}

\begin{center}
\textrm{
\newline
\newline
\newline
\newline
\newline
\newline
\newline
R. Elliott et. al, Fig. 1 
}
\end{center}

\clearpage

\clearpage

\begin{figure}[!t]
\includegraphics[scale=0.80]{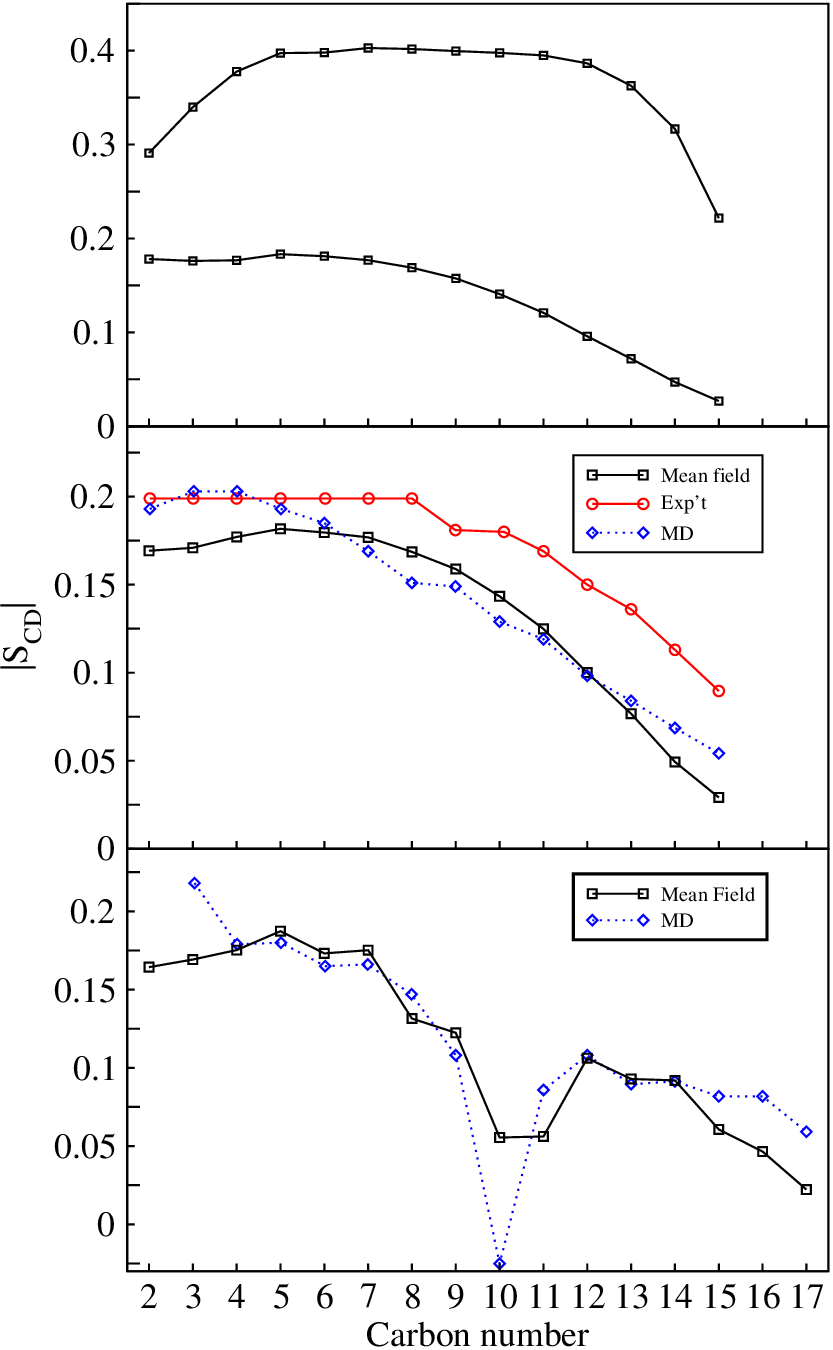}
\label{dppc_order}
\end{figure}

\begin{center}
\textrm{
\newline
\newline
\newline
\newline
R. Elliott et. al, Fig. 2 
}
\end{center}

\clearpage

\clearpage
\begin{figure}[!t]
\includegraphics[scale=0.70]{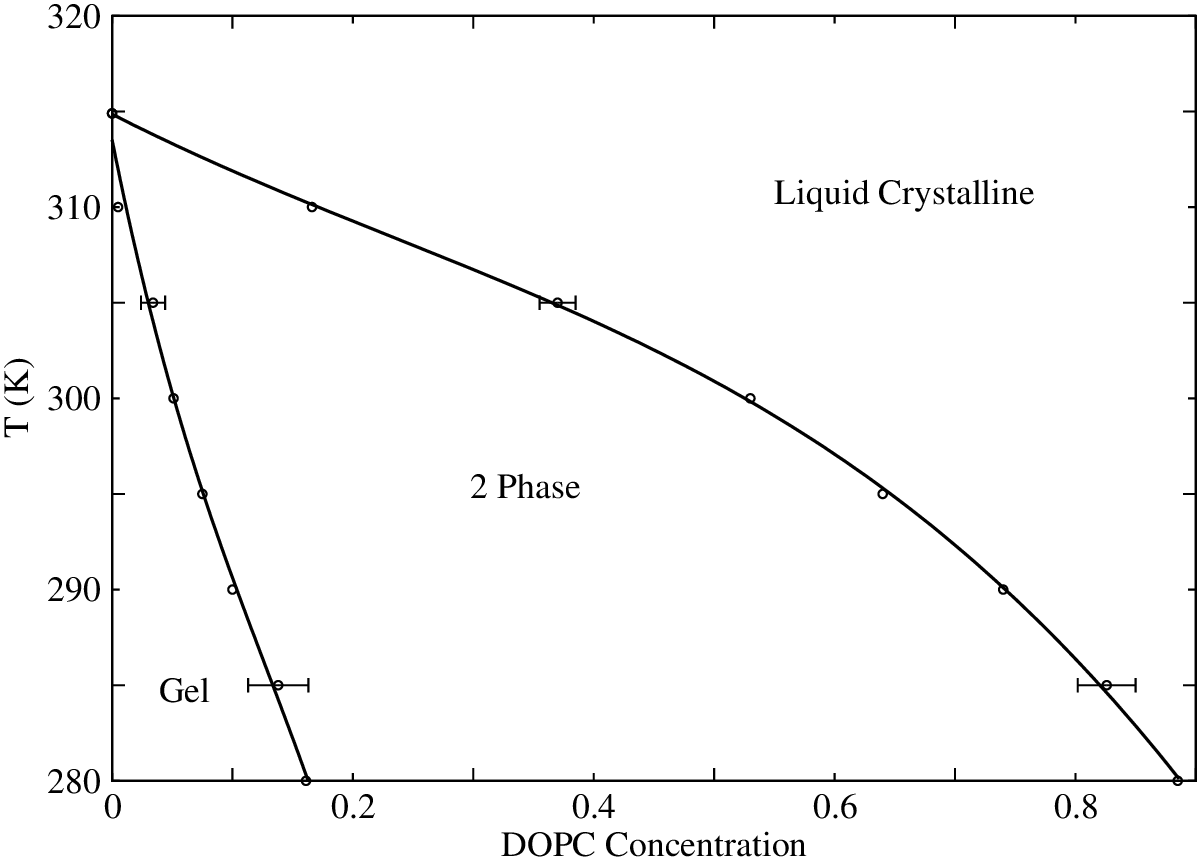}
\label{big_result}
\end{figure}

\begin{center}
\textrm{
\newline
\newline
\newline
\newline
\newline
\newline
\newline
R. Elliott et. al, Fig. 3 
}
\end{center}

\clearpage

\clearpage

\begin{figure}[!t]
\includegraphics[scale=0.75]{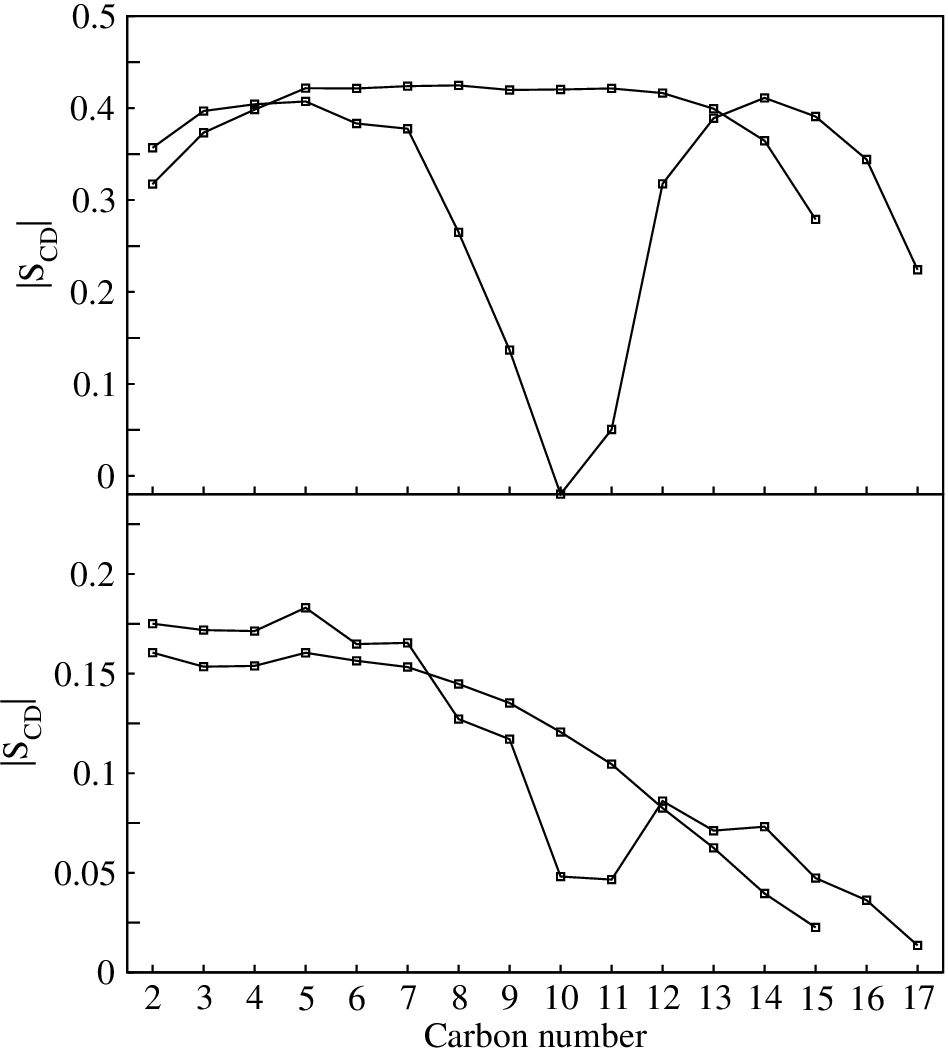}

\label{both_scd}
\end{figure}

\begin{center}
\textrm{
\newline
\newline
\newline
\newline
\newline
\newline
\newline
R. Elliott et. al, Fig. 4 
}
\end{center}

\end{document}